\documentclass[%
 reprint,
superscriptaddress,
 amsmath,amssymb,
 aps,
floatfix,
]{revtex4-1}
\usepackage{graphicx}
\usepackage{dcolumn}
\usepackage{bm}

\usepackage{dsfont}
\usepackage{verbatim}
\usepackage{xcolor}

\definecolor{darkgreen}{rgb}{0.0, 0.5, 0.0}

\newcommand{\EE}{Department of Electrical Engineering, Princeton University, Princeton, New Jersey 08540, USA}
\newcommand{\NIST}{Material Measurement Laboratory, National Institute of Standards and Technology, Gaithersburg, Maryland 20899, USA}
\newcommand{\CFN}{Center for Functional Nanomaterials, Brookhaven National Laboratory, Upton, New York 11973, USA}
\newcommand{\Ph}{Department of Physics, Princeton University, Princeton, New Jersey 08540, USA}
\newcommand{\NSLS}{National Synchrotron Light Source II, Brookhaven National Laboratory, Upton, New York 11973, USA}
\newcommand{\Ang}{Angstrom Engineering Inc. Kitchener, Ontario, Canada N2E1W8}

\begin{document}

\title{Microscopic Relaxation Channels in Materials for Superconducting Qubits}

\author{Anjali Premkumar}
\email[Corresponding author: ]{anjalip@princeton.edu}
\affiliation{\EE}
\author{Conan Weiland}
\affiliation{\NIST}
\author{Sooyeon Hwang}
\affiliation{\CFN}
\author{Berthold J\"{a}ck}
\affiliation{\Ph}
\author{Alexander P. M. Place}
\affiliation{\EE}
\author{Iradwikanari Waluyo}
\affiliation{\NSLS}
\author{Adrian Hunt}
\affiliation{\NSLS}
\author{Valentina Bisogni}
\affiliation{\NSLS}
\author{Jonathan Pelliciari}
\affiliation{\NSLS}
\author{Andi Barbour}
\affiliation{\NSLS}
\author{Mike S. Miller}
\affiliation{\Ang}
\author{Paola Russo}
\affiliation{\Ang}
\author{Fernando Camino}
\affiliation{\CFN}
\author{Kim Kisslinger}
\affiliation{\CFN}
\author{Xiao Tong}
\affiliation{\CFN}
\author{Mark S. Hybertsen}
\affiliation{\CFN}
\author{Andrew A. Houck}
\email[Corresponding author: ]{aahouck@princeton.edu}
\affiliation{\EE}
\author{Ignace Jarrige}
\email[Corresponding author: ]{jarrige@bnl.gov}
\affiliation{\NSLS}

\begin{abstract}
Despite mounting evidence that materials imperfections are a major obstacle to practical applications of superconducting qubits, connections between microscopic material properties and qubit coherence are poorly understood. Here, we perform measurements of transmon qubit relaxation times $T_{1}$ in parallel with spectroscopy and microscopy of the thin polycrystalline niobium films used in qubit fabrication. By comparing results for films deposited using three techniques, we reveal correlations between $T_{1}$ and grain size, enhanced oxygen diffusion along grain boundaries, and the concentration of suboxides near the surface. Physical mechanisms connect these microscopic properties to residual surface resistance and $T_1$ through losses arising from the grain boundaries and from defects in the suboxides. Further, experiments show that the residual resistance ratio can be used as a figure of merit for qubit lifetime. This comprehensive approach to understanding qubit decoherence charts a pathway for materials-driven improvements of superconducting qubit performance.

\end{abstract}

\maketitle

\thispagestyle{empty}

\section*{Introduction}
Following two decades of dramatic improvement, superconducting qubit technology has emerged as a promising platform for fault-tolerant quantum computation. Game-changing enhancements in coherence have especially been achieved through novel designs and improved fabrication processes. However, performance improvements by these means have started to plateau, indicating that the dominant sources of decoherence are not well understood \cite{Kjaergaard2019}. This has led to a recent surge of interest in understanding the limiting loss mechanisms in qubit materials \cite{Wang2015,Dial2016,Gambetta2017,Burnett2019,Schlor2019,Nersisyan2019, Woods2019}. 

In particular, recent breakthroughs have highlighted the central role played by uncontrolled surfaces and interfaces in the decoherence of transmon qubits \cite{Oliver2013, Martinis2014,Burnett2019,Schlor2019,Nersisyan2019}. Proposed mechanisms for loss at surfaces and interfaces involve interactions between the qubit and microscopic objects such as impurity-based paramagnetic defects \cite{Lee2014}, charge-trapping defect sites \cite{VanHarligen2004}, and two-level systems (TLS) \cite{Simmonds2004,Lisenfeld2010,Oliver2013,Martinis2014,Paz2014,Muller2019,Burnett2019,Schlor2019} characterized by atoms or charges tunneling between two metastable states. Several studies have quantified the macroscopic impact of such interactions by showing a proportionality between relaxation and participation ratios at various interfaces \cite{Wang2015,Dial2016,Gambetta2017,Woods2019}. Other studies have mapped interactions between the qubit and individual TLS's to fluctuations in relaxation, dephasing, and qubit frequency \cite{Burnett2019, Schlor2019}. However, plausible identifications of specific defects or impurities associated with such phenomenological findings are rare. To address this challenge, a substantial scope of multidisciplinary research is required to explore the large phase space of potentially relevant materials properties and identify connections with qubit performance.

\begin{figure*}[t]
\centering
\includegraphics[scale=1]{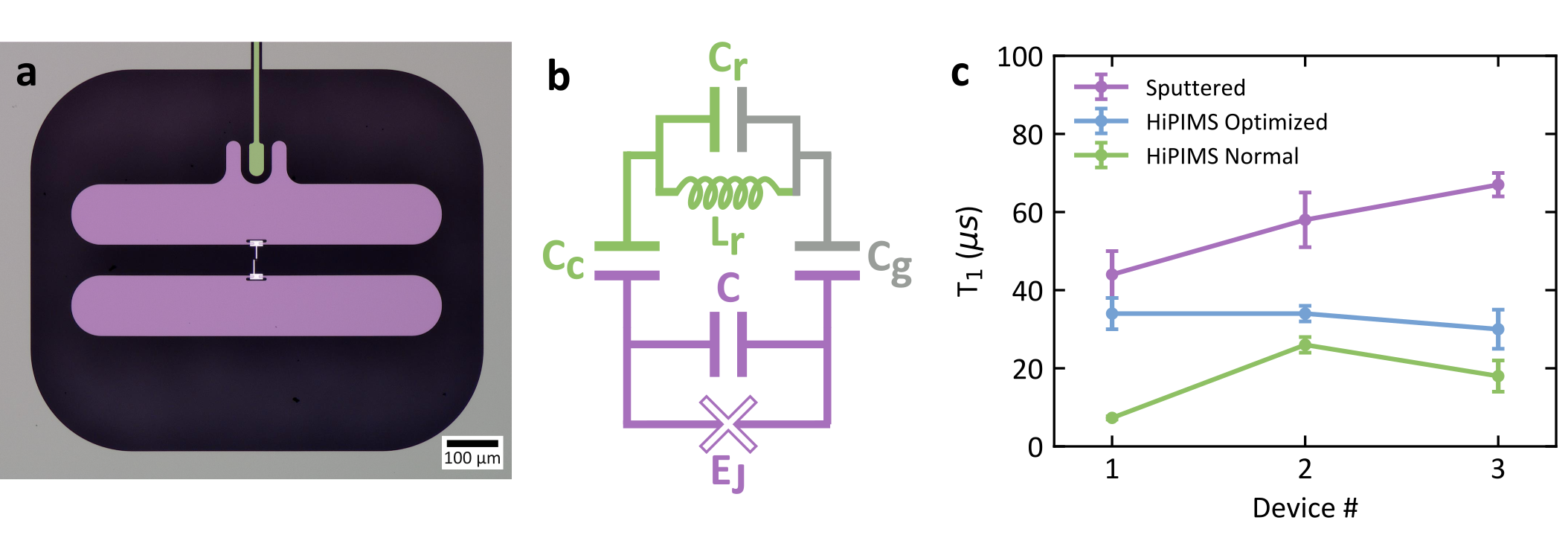}
\caption{Transmon qubit design and performance. (a) False color optical image of a representative transmon qubit from our study. Niobium regions include the center-pin of the coplanar waveguide resonator (green), the transmon capacitor pads (purple), and the ground plane (gray). The aluminum Josephson junction is shown in white. Black areas indicate where the metal has been etched away, and the sapphire substrate is exposed. (b) Effective circuit diagram of a transmon qubit coupled to a resonator. Each circuit element is schematically colored as in (a). The resonator is comprised of a center-pin coupled to ground via a capacitor ($C_r$) and an inductor ($L_r$). $E_J$ and $C$ refer to the Josephson energy and the capacitance of the qubit. The qubit is capacitively coupled to the center-pin of the resonator ($C_c$) and to ground ($C_g$). (c) Measured relaxation times ($T_1$) for three rounds of devices fabricated with sputtered, HiPIMS optimized, and HiPIMS normal niobium films. Error bars indicate standard deviation.}
\label{fig:qubit}
\end{figure*}

In this study, we bridge the gap between qubit performance and microscopic materials properties through a combined materials-and-device investigation of transmon qubits. Using spatially resolved x-ray spectroscopy and microscopy, we characterized the structural and electronic properties of the niobium (Nb) thin films used to fabricate transmon qubit devices. We compared three different deposition methods for the Nb films, yielding a quantifiable variation in nanostructure, surface oxide composition, and transport properties. Several of these film properties correlate with the measured qubit relaxation time $T_1$. Specifically, for qubits with shorter relaxation times, the corresponding films exhibited smaller grain size, larger voids at grain boundaries, enhanced oxygen diffusion along grain boundaries near the surface, and excess suboxides at the interface between the Nb metal and the Nb$_{2}$O$_{5}$ passivation layer. Additionally, we observed a correlation between shorter relaxation times and higher residual resistance in the Nb films. We discuss mechanisms that link the observed microscopic features to residual resistance and $T_1$, specifically dissipation at grain boundaries and defect-induced dielectric loss from suboxides. These results are a critical first step in connecting precise materials properties with microscopic models for decoherence and establishing a materials-based approach to enhance qubit performance.

\section*{Results}
\textbf{Qubit Design and Performance}. Qubit characterization was performed on transmon qubits, which are widely used for quantum computing and quantum simulation applications \cite{Koch2007,Arute2019,Kjaergaard2019,Ma2019,Mundada2019,Burnett2019,Schlor2019,Guo2019}. Transmons consist of a Josephson junction---implemented with a thin aluminum oxide barrier between superconducting wires---shunted by a large capacitor (Fig. \ref{fig:qubit}a). The Josephson junction acts as a non-linear inductor, creating an anharmonic oscillator where the two lowest-lying states can be used as a coherent qubit \cite{Koch2007}. Transmons are controlled and measured in a circuit quantum electrodynamics (cQED) platform, where the qubit is capacitively coupled to a coplanar waveguide resonator and operated in the dispersive regime (Fig. \ref{fig:qubit}b). Measurements are performed by monitoring transmission at the resonator frequency, which shifts as a function of qubit state \cite{Blais2004}.

We employ three different deposition methods for the Nb films used to fabricate the coplanar waveguide resonators, qubit capacitor pads, and ground plane in the transmon devices. All films are deposited on sapphire substrates. First, we use DC sputter deposition, which is common in superconducting qubit fabrication. We additionally study two variations of high-power impulse magnetron sputtering (HiPIMS), where the DC ionization voltage is replaced with short, high-power pulses, potentially leading to a higher degree of ionization and denser films \cite{Alami2006, Kouznetsov1999, Soucek2017}. The average cathode power in HiPIMS is typically chosen to be similar to standard DC sputtering; such a process was performed on our first HiPIMS variation, which we refer to as ``HiPIMS normal". Our second variation, ``HiPIMS optimized", involves an optimized geometry of the Nb target and sapphire substrate and a higher peak current, resulting in triple the average power.

We performed relaxation ($T_1$) measurements to characterize the dependence of qubit performance on deposition technique. Since $T_1$ is limited by microwave losses in the superconductors and substrate, it is a valuable probe of materials-induced loss channels. We measured three rounds of devices, where a given round consisted of one device made with each Nb deposition technique, to maintain as consistent fabrication, packaging, and shielding as possible within each round. Details of qubit fabrication, measurement setup, and qubit performance are in the Supplementary Information. The measured $T_1$ times for each film type show a clear statistical separation between the three deposition techniques (Fig. \ref{fig:qubit}c). The sputtered niobium qubit consistently performs the best, followed by the HiPIMS optimized qubit, then the HiPIMS normal qubit. The average qubit relaxation times are also displayed in Table \ref{table:summary}.  We employ a wide range of characterization techniques to understand possible microscopic origins of these coherence differences.

\begin{figure*}[ht]
\centering
\includegraphics[scale=0.7]{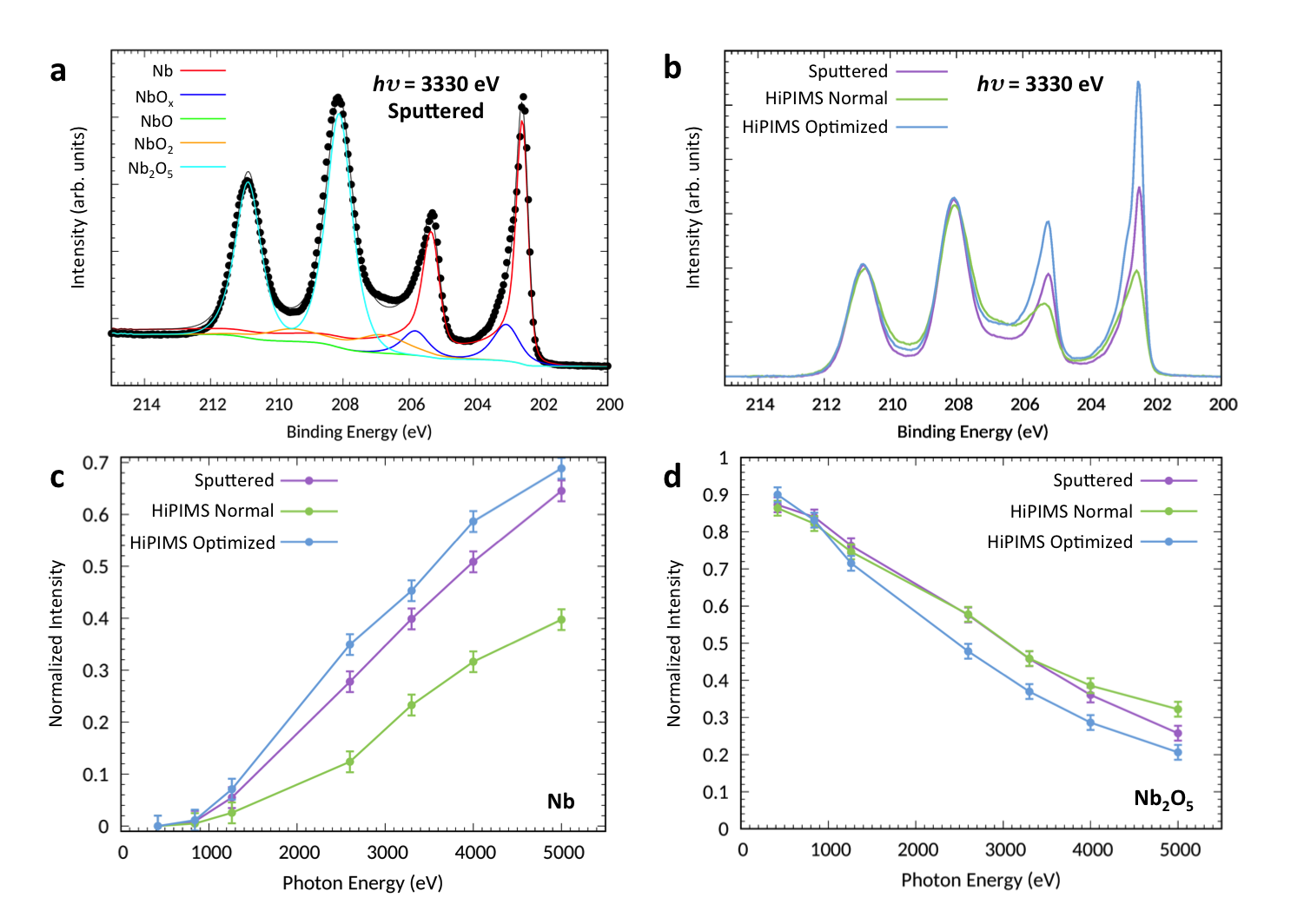}
\caption{(a) Representative x-ray photoemission spectroscopy (PES) spectrum of the Nb $3d_{3/2}$ and $3d_{5/2}$ core levels, measured on the sputtered film for a photon energy of 3330 eV (black dots) and fitted with five components. (b) Measured spectra for all three film types at 3330 eV, normalized to the intensity of the Nb$_{2}$O$_{5}$ component. For each film, the measured intensity of the Nb (c) and Nb$_{2}$O$_{5}$ (d) peaks are plotted at several photon energies. The sum of the signals from the different oxidation states in a given film is normalized to one, and the error bars show a 1 \% error, as estimated from the signal-to-noise of measured data. The intensity of Nb and Nb$_{2}$O$_{5}$ increase and decrease with energy, respectively, indicating the presence of a surface oxide layer.}
\label{fig:pes1}
\end{figure*}
\

\begin{figure*}[ht]
\centering
\includegraphics[width=\linewidth]{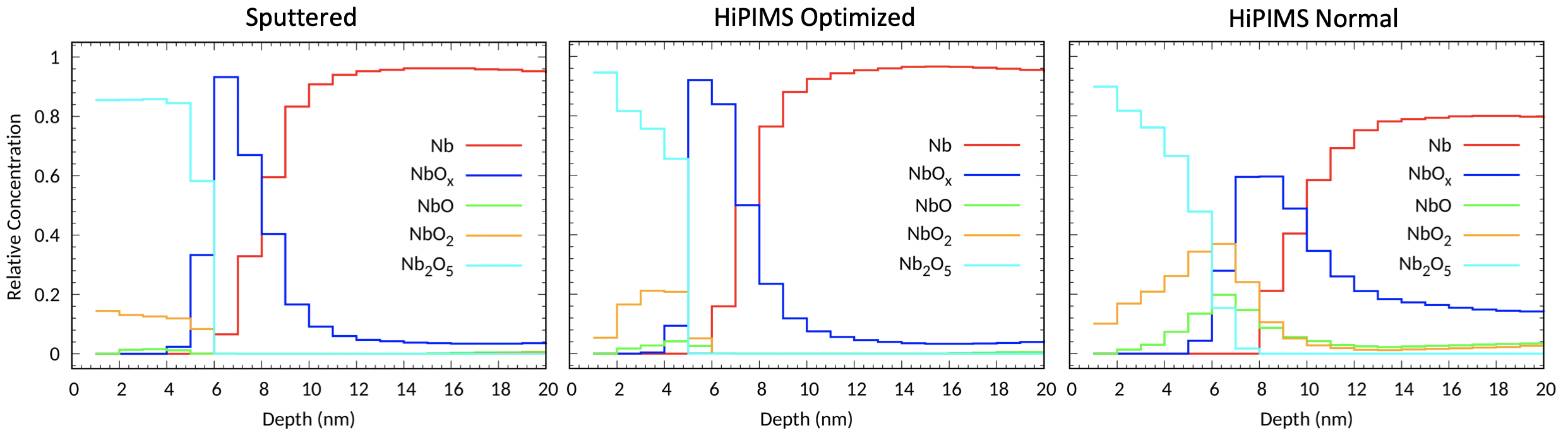}
\caption{Depth profiles of the different oxidation states of Nb reconstructed from PES data using a maximum-entropy method (MEM) algorithm for the three types of Nb film. Each film shows a surface layer of a few nm of Nb$_{2}$O$_{5}$, a transition layer with varying concentrations of different suboxides, and the Nb metal bulk. In particular, the HiPIMS normal film shows significant concentrations of NbO and NbO$_2$ in the transition layer and deeper penetration of NbO$_x$ into the metal.}
\label{fig:pes2}
\end{figure*}

\begin{figure}[ht]
\centering
\includegraphics[scale=0.47]{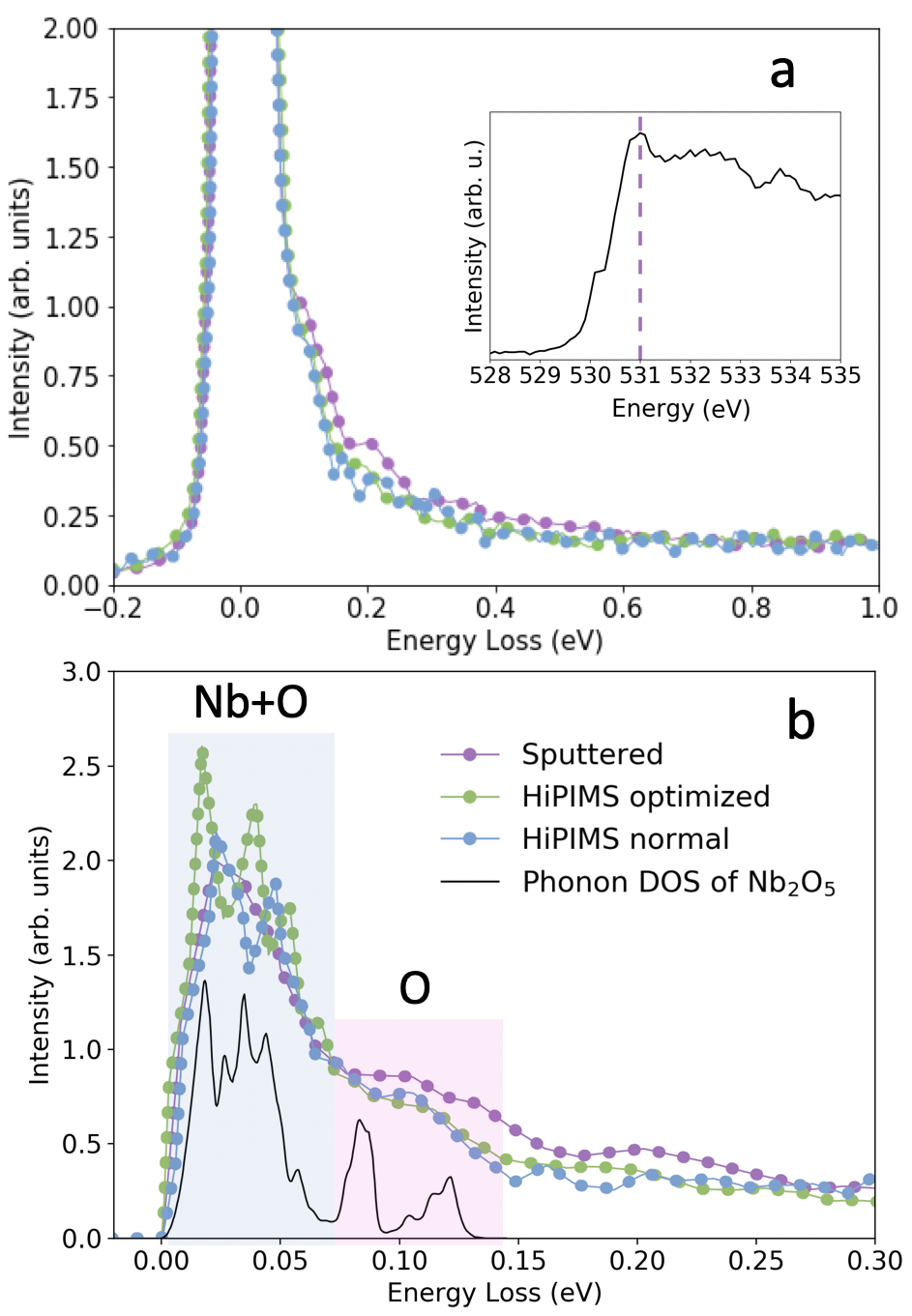}
\caption{(a) Resonant inelastic X-ray scattering (RIXS) spectra measured for the three types of films at the oxygen $K$-edge resonance, for an incident energy of 531 eV. The inset shows the O-$K$ absorption spectrum of the sputtered film with a vertical dashed line at the resonance. (b) Close-up view of the RIXS spectra after subtraction of the elastic line, with the phonon density of states (DOS) calculated for Nb$_{2}$O$_{5}$ from \cite{Cheng2019}. The overall scaling factor of the DOS was chosen to aid visualization. The DOS was reported to arise from both Nb and O up to $\approx$70 meV, and mostly from O at higher energies, as represented by the blue and pink bands, respectively.}
\label{fig:rixs}
\end{figure}

\begin{figure*}[p!]
\centering
\includegraphics[scale=0.57]{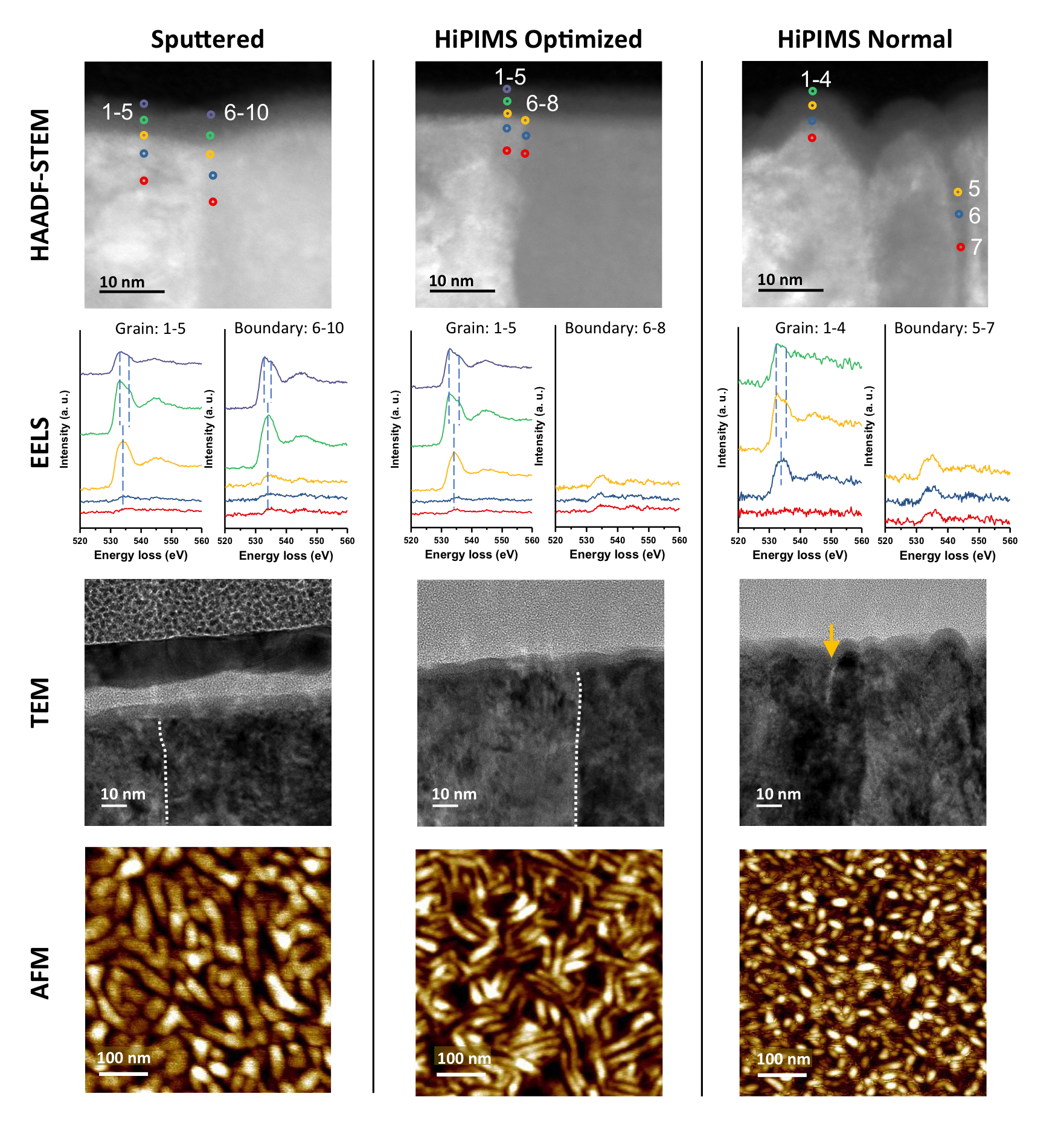}
\caption{Structural and chemical imaging of the sputtered (left), HiPIMS optimized (middle), and HiPIMS normal (right) films. The top row shows high-angle annular dark-field scanning transmission electron microscopy (HAADF-STEM) measurements at cross sections of the films' surfaces, revealing a $\approx$5 nm oxide layer and variations in grain size. The second row shows O-$K$-edge electron-energy loss spectroscopy (EELS) spectra measured at the locations indicated on the HAADF-STEM images. For the sputtered and HiPIMS optimized films, both the EELS spectra within a grain (left) and the spectra taken along a grain boundary (right) show a transition from a double peak (Nb$_{2}$O$_{5}$) to a single peak (suboxides) to a negligible peak (metal). However, for the HiPIMS normal film, EELS spectra along the grain boundary reveal similar oxidation peaks to the surface oxide layer, indicating that oxygen has diffused into the grain boundary to form oxides. The third row shows TEM bright-field images of cross sections of the films' surfaces, where the white dotted lines delineate grain boundaries for the sputtered and HiPIMS optimized films, and the yellow arrow points to a gap at the grain boundary for the HiPIMS normal film. The grainy, light gray layer above the surface is platinum, which protects the surface during sample preparation. The bottom row shows atomic force microscopy (AFM) images measured over a 500 nm x 500 nm area. It is visually evident that the sputtered film grain size is the largest, and the HiPIMS normal film grain size is the smallest.}
\label{fig:tem-eels}
\end{figure*}

\textbf{Surface oxides}. The surface oxides on the three types of Nb films were investigated using a combination of soft and hard X-ray photoemission spectroscopy (PES) and resonant inelastic X-ray scattering (RIXS). The experimental details can be found in the Methods section. Figure \ref{fig:pes1}a shows a representative spectrum of the spin-orbit-split Nb $3d_{3/2}$ and $3d_{5/2}$ core levels, measured on the sputtered film with an incident energy of 3330 eV. Also shown are the fits to distinct materials components. The PES spectra are best fit with five doublet components, revealing that there are four coexisting oxidation states besides Nb metal: NbO$_{x}$ (x$<$1), NbO, NbO$_{2}$, and Nb$_{2}$O$_{5}$, with respective shifts in binding energy of +0.5 eV, +1.8 eV, +3.8 eV, and +5.6 eV compared to Nb metal. The presence of five total oxidation states including Nb metal is consistent with previous PES studies \cite{Tian2008}. Figure \ref{fig:pes1}b shows the spectra measured on all three types of film at 3330 eV, revealing clear variations in the concentrations of different oxidation states among the different deposition techniques.

PES measurements of the Nb $3d$ core levels were performed for seven values of the incident photon energy between 420 eV and 5000 eV. Since higher incident photon energies probe deeper into the film, this energy range resulted in surface sensitivity from the topmost monolayers down to a depth of $\approx$20 nm. The relative intensities of the signals from the five oxidation states were extracted for different photon energies, as shown for Nb metal and Nb$_{2}$O$_{5}$ in Figure \ref{fig:pes1}c and \ref{fig:pes1}d respectively. As expected, the Nb and Nb$_{2}$O$_{5}$ components respectively increase and decrease with increasing photon energy, which is consistent with the presence of a surface oxide layer. To reconstruct a depth profile of the concentrations of the oxides, we applied a maximum-entropy method (MEM) algorithm. The MEM is a statistical method used to obtain the most likely depth profile with minimal user input from the energy-dependent fits to the spectra. It has been commonly applied to angle-resolved PES \cite{Chang2000} and, more recently, to variable photon energy PES \cite{Weiland2014, Weiland2016} as in the present study.

The resulting depth profiles of the relative concentrations of the different oxidation states are shown in Figure \ref{fig:pes2}. For all three film types, we observe a surface oxide layer with Nb$_{2}$O$_{5}$ as the main constituent, followed by a transition layer of lower oxidation states whose thickness, depth distribution, and composition vary between the different films. The oxide/metal interface is sharpest for the sputtered film and is increasingly diffuse for the HiPIMS optimized and then the HiPIMS normal film. The HiPIMS optimized film shows some Nb$_{2}$O$_{5}$-NbO$_{2}$ interdiffusion in the 2 nm to 6 nm range, as seen from the respectively decreasing and increasing profiles of Nb$_{2}$O$_{5}$ and NbO$_{2}$, and it also shows a thicker NbO$_{x}$ layer than the sputtered film. The Nb$_{2}$O$_{5}$-NbO$_{2}$ interdiffusion is strongest in the HiPIMS normal film, and the formation of NbO is also evident around 6 nm. Significantly, the NbO$_{x}$ penetrates most deeply in the HiPIMS normal film, extending beyond the 20 nm depth sensitivity allowed by this technique. The relative concentrations of the different Nb oxidation states, estimated by integrating over the 20 nm depth profiles and expressed in percents, are displayed in Table \ref{table:summary}.

RIXS is a complementary probe of the electronic structure that offers sensitivity to low-energy excitations of the electronic and lattice degrees of freedom (phonons) \cite{Ament2011}. RIXS spectra were measured on all three films at room temperature and at a grazing incident angle of $10^{\circ}$, yielding a depth sensitivity of $\approx$10 nm. Measurements were performed at the oxygen $K$-edge (O-$K$), a band edge corresponding to excitations from the oxygen 1s core level into unoccupied valence states, thereby providing chemical sensitivity to the oxide layer. The spectra in Figure \ref{fig:rixs}a were collected for an incident photon energy tuned to the resonance of the O-$K$ absorption, shown for the sputtered film in the inset. They are normalized in intensity to the area under the peak of the emission line (not shown), which spans the $\approx$3 eV to 11 eV energy loss range.

Figure \ref{fig:rixs}b shows a close-up view of the low-energy features after removing the elastic peak, achieved by subtracting the negative-energy-loss tail mirrored about zero energy. The energy range and broad profile of the inelastic features are reminiscent of a multi-phonon shake-up process \cite{Bisogni2012, Lee2013}, specifically phonons in the surface oxides, dictated by the use of the O-$K$-edge resonance at grazing incidence. The inelastic features are similar in energy and lineshape across the three films, with a slightly higher intensity in the 0.1 eV to 0.3 eV energy loss range for the sputtered film. These spectra are in stark contrast to the multiple vibrational modes extending up to $\approx$ 2 eV that would be expected from OH groups \cite{Ertan2017}, which indicates that there are no detectable hydrogen bonds in the surface oxides of these films. This finding, which is consistent with previous reports that the surface Nb oxide layer acts as a barrier against H diffusion \cite{Isagawa1980}, is significant as hydrogen bonds in Al$_{2}$O$_{3}$ have been proposed as one of the primary sources of TLS resonant absorption in superconducting qubits \cite{Jameson2011,Gordon2015,deGraaf2017}. This suggests that TLS's from hydrogen bonds in particular are unlikely to be a dominant source of loss in our Nb films.

Figure \ref{fig:rixs}b also shows the phonon density of states (DOS) calculated using density functional theory, taken from ref. \cite{Cheng2019}. The main features of the DOS are consistent with the energy and relative intensities of the experimental peaks. Ref. \cite{Cheng2019} showed that both Nb and O vibrations contribute to the features below 70 meV (blue region), whereas the higher-energy phonons mostly arise from O because of its smaller atomic mass (pink region). The two HiPIMS films show a lower intensity at higher energies, indicating a larger quantity of O vacancies in the Nb$_{2}$O$_{5}$ layer. This lower intensity cannot be explained by a lower oxidation state, as the theoretical phonon DOS of suboxides have a stronger, and not weaker, relative intensity for the high-energy features compared with Nb$_{2}$O$_{5}$, as shown for NbO$_{2}$ in Ref. \cite{O'Hara2015}.

\setlength{\tabcolsep}{6pt}
\renewcommand{\arraystretch}{1.8}
\setlength{\arrayrulewidth}{0.3mm}
\begin{table}[t]
\centering
\begin{tabular}{c|c c c}
Deposition & Sputtered & HiPIMS opt & HiPIMS norm \\
\hline
T$_{1}$ ($\mu$s) & 56 $\pm$ 12 & 33 $\pm$ 2 &  17 $\pm$ 9\\
RRR & 8.9 $\pm$ 0.1 & 5.0 $\pm$ 0.2 & 2.9 $\pm$ 0.1\\
T$_\mathrm{c}$ (K) & 9.0 $\pm$ 0.1 & 8.6 $\pm$ 0.1 & 8.1 $\pm$ 0.1\\
GSA (nm$^{2}$) & 1140 $\pm$ 70 & 500 $\pm$ 50 & 180 $\pm$ 30\\
Nb & 61 $\pm$ 3 & 64 $\pm$ 3 & 45 $\pm$ 2\\
NbO$_{x}$ & 15.1 $\pm$ 0.2 & 16 $\pm$ 0.3 & 20.4 $\pm$ 0.8\\
NbO & 0 $\pm$ 2 & 0 $\pm$ 1 & 5 $\pm$ 1\\
NbO$_{2}$ & 3.1 $\pm$ 0.4 & 3.5 $\pm$ 0.2 & 10 $\pm$ 2\\
Nb$_{2}$O$_{5}$ & 20 $\pm$ 1 & 15.9 $\pm$ 0.8 & 19 $\pm$ 2\\
Suboxide & 19 $\pm$ 2 & 20 $\pm$ 1 & 36 $\pm$ 2\\
\end{tabular}
\caption{Summary of the main parameters extracted from qubit and thin film measurements. For $T_1$, we report the average and standard deviation between the three rounds for each film. GSA is the mean grain surface area. The numbers for the different oxidation states of Nb are relative concentrations integrated over the 20 nm in the depth profiles of Figure \ref{fig:pes2}, expressed in percentages. The errors were calculated for each depth step, and summed in quadrature and weighted according to the total measured signal for each oxidation state. Suboxide is the sum of the relative concentrations of NbO$_{x}$, NbO, and NbO$_{2}$.}
\label{table:summary}
\end{table}

\textbf{Structural and chemical imaging}. In order to correlate our surface oxide spectroscopy findings with morphology and grain size, we performed transmission electron microscopy (TEM), spatially resolved electron-energy loss spectroscopy (EELS), and atomic-force microscopy (AFM) measurements on all three types of Nb films. The results are summarized in Figure \ref{fig:tem-eels}. In the top row, we present high-angle annular dark-field scanning transmission electron microscopy (HAADF-STEM) images of film cross sections at the surface. This measurement confirms the presence of a $\approx$5 nm thick oxide layer at the surface of all three films. The near-surface morphology of the HiPIMS normal film is visibly different, with the oxide layer conforming to smaller grains.

To characterize local chemical properties near the surface, we performed spatially resolved EELS. The second row shows O-$K$-edge EELS spectra that were measured at the locations indicated by colored dots on the HAADF-STEM images. In each EELS panel, the left plot shows chemical variation from the surface into the grain, and the right plot shows variation along grain boundaries. The main O-$K$-edge feature near 534~eV shows a transition from a clear double-peak line shape near the surface to a single broad peak near the interface with Nb metal. Based on EELS spectra previously reported for different Nb reference oxides \cite{Bach2009}, this indicates a change from Nb$_{2}$O$_{5}$ to lower oxidation states, in agreement with the PES measurements. For the sputtered and HiPIMS optimized films, spectra along grain boundaries (locations 8-10 and 6-8, respectively) do not reveal a stronger oxidation than in the Nb metal away from the grain boundaries (locations 4 and 5 in both films). However, for the HiPIMS normal film, the presence of the oxide in the boundary is clear, as seen from the stronger signal in locations 5-7 compared with location 4. These results are consistent with the overall picture from PES profiles shown in Figure \ref{fig:pes2}, which show significant oxygen diffusion for the HiPIMS normal films. By contrast, further TEM-EELS analysis reveals that for all films, the substrate-metal interface is chemically sharp, with no oxides observed at grain boundaries (See Supplementary Figures S5 and S6).

To further characterize grain boundaries, bright-field TEM images that rely on a diffraction contrast from the transmitted beam were taken at cross sections of each sample, as shown in the third row of Figure \ref{fig:tem-eels}. Grain boundaries, indicated by white dotted lines for the sputtered and HiPIMS optimized films, are observed in each case. The grain structure observed using TEM appears to be largely columnar, with grains extending through the thickness of the films (see Supplementary Figure S4). The HiPIMS normal film is distinguished by smaller grain sizes, and it exhibits a lower packing density with apparent gaps along some of the grain boundaries (yellow arrow). The EELS measurements suggest that these gaps host oxidized Nb near the film surface. 

Further information about the grain morphology and size can be extracted from the complementary, top-down view provided by the AFM images shown in the bottom row of Figure \ref{fig:tem-eels}. First, it appears that the grains are elongated and not spherical. Second, the grain size decreases between the sputtered and HiPIMS optimized films, and even more significantly between the HiPIMS optimized and normal films. This observation is quantified by estimating the mean grain surface areas using a watershed algorithm \cite{Rabbani2015}, as reported in Table \ref{table:summary}. We also used AFM to measure short-range film roughness, which is found to be approximately 1 nm RMS for all three films.

\textbf{Residual resistance ratio}. Measured values of residual resistance ratio (RRR) for all three film types are displayed in Table \ref{table:summary}. RRR is defined as the ratio between the values of the resistivity near room temperature and just above the superconducting critical temperature $T_\mathrm{c}$, and it quantifies the impact of defect scattering on bulk DC conductivity. We specifically report RRR values as the ratio of measured resistivity between 295 K and 10 K. The sputtered film has the highest RRR, followed by HiPIMS optimized, and then HiPIMS normal. We also report measured $T_\mathrm{c}$, which shows a similar trend.

\section*{Discussion}
We observe correlations between $T_1$, RRR, grain size, and suboxide concentration near the surface, as shown by the data in Table \ref{table:summary}. RRR has long been used as a gauge of the quality of superconducting Nb films for the construction of superconducting radio-frequency (RF) cavities \cite{Russo2005, Valderrama2012}. Further, seminal studies of RF cavities with Nb walls have shown that residual surface resistance limits the cavity quality factor at low temperatures \cite{Turneaure1968, Broom1977, Halbritter1971, Schmuser2002, Romanenko2017}. Thus, our finding that $T_1$ correlates with RRR suggests that the relaxation time may be limited by the intrinsic qubit quality factor arising from losses in the Nb films. In the following, we connect our results to plausible mechanisms by which grain boundaries and suboxides lead to residual surface resistance and thus limit $T_1$.

First, we discuss how smaller grain size enhances microwave surface resistance. The correlation we observe between grain size, RRR, and $T_\mathrm{c}$ is consistent with a body of literature that links these quantities in polycrystalline Nb thin films \cite{Bose2005,Bose2006a,Bose2006b}. The relationship between grain size and RF surface resistance in Nb films has been understood by describing grain boundaries as Josephson weak-links \cite{Bonin1991, Attanasio1991}. Under the assumption that the grain boundaries' critical current and shunt resistance are independent of grain size, the surface resistance simply depends on grain boundary density, and thus scales inversely with the grain size \cite{Bonin1991}. Combined with our finding that residual resistance scales inversely with $T_1$, this scaling is consistent with the measured relaxation times and grain sizes across the three deposition techniques. Confirmation of the Josephson weak-link model hinges on further characterization of the grain boundaries and their electrical characteristics. These characteristics likely vary across our three film types based on measured differences in grain boundary composition, as we discuss next.

The HiPIMS normal films not only exhibit smaller grain size, but also a lower packing density, leading to significant grain boundary voids which host oxides near the surface (Fig. \ref{fig:tem-eels}). This oxide is in fact NbO$_{x}$, which penetrates into the metal (Fig. \ref{fig:pes2}). Since NbO$_x$ forms through the migration of interstitial oxygen defects resulting in local lattice distortions \cite{Dosch1986,Delheusy2008}, its presence in the grain boundaries can contribute to microwave absorption, including through TLS. Given that the effective penetration depth of Nb at 6 GHz is on the order of 60 nm at low temperatures \cite{Gao2008}, NbO$_{x}$ extending over 20 nm into the film can have a sizeable impact. By contrast, in the sputtered and HiPIMS optimized films, NbO$_{x}$ penetrates only a few nanometers into the metal (Fig. \ref{fig:pes2}), and oxides are undetectable in the grain boundaries using EELS (Fig. \ref{fig:tem-eels}), suggesting relatively less impact on microwave absorption in these films. 

Another potential dissipation mechanism involves the quality of the surface oxide layer. The Nb$_{2}$O$_{5}$ passivation layer is found to have the lowest concentration of oxygen vacancies and to be most homogeneous in the sputtered film (Figs. \ref{fig:rixs} and \ref{fig:pes2}, respectively). The Nb$_{2}$O$_{5}$ layer is least homogeneous in the HiPIMS normal film, where NbO$_{2}$ and NbO are observed at the interface with Nb. The presence of different oxidation states is expected to result in a more complex oxide-metal interface, where mismatches in crystalline parameters lead to increased strain and dislocation density. NbO itself is a source of defects, as it crystallizes in a naturally defective structure with a quarter of the Nb and O sites vacant. These sources of elevated defect density can enhance dielectric loss through contributions from paramagnetic carriers \cite{Lee2014} and bistable structural motifs, which have been suggested to host TLS \cite{Paz2014,Muller2019}. Additionally, while NbO is a superconductor, NbO$_{2}$ is not, further contributing to loss. In Table \ref{table:summary}, we report the sum of the relative concentrations of the suboxides NbO$_{x}$, NbO and NbO$_{2}$, which exhibits a negative correlation with $T_{1}$. This observation indicates that the total suboxide concentration potentially limits relaxation times in superconducting qubits. 

In conclusion, this study builds a significant bridge between the performance of superconducting transmon qubits and the materials properties of Nb films, a critical component in qubit fabrication. Our cross-cutting investigation exploits microscopic variations among Nb thin films deposited using three sputtering techniques, and reveals correlations between qubit relaxation time and structural and chemical Nb film properties, specifically grain size, suboxide intergranular penetration, and suboxide intragrain concentration near the surface. Significantly, the measured RRR serves as a proxy for the observed variation in qubit relaxation time and these film properties. We anticipate that further improvements in relaxation times will arise from optimizations of grain size in Nb deposition techniques and from exploring the use of superconductors that show a more chemically homogeneous surface oxide layer than Nb, such as tantalum \cite{Place2020}. More broadly, the connections between material properties and relaxation mechanisms uncovered in this work form a solid basis for physical models to guide the development of materials for superconducting qubits. Future multidisciplinary efforts can also leverage the multimodal spectroscopy and microscopy approach used here to make further progress toward identifying microscopic mechanisms for decoherence and engineering high-performance qubits.

\section*{Methods}
\subsection*{Materials characterization}

Synchrotron measurements were carried out at the National Synchrotron Light Source II (NSLS-II) \cite{Hill2020}, in Brookhaven National Laboratory (BNL). Hard X-ray photoemission spectroscopy (HAXPES) measurements were performed on beamline 7-ID-2 of the National Institute of Standards and Technology (NIST) using a 400 mm diameter hemispherical electron energy analyzer with a 300 um slit, mounted at 90 degrees with respect to the beam propagation, and operating at 200 eV pass energy. A double crystal monochromator was used for x-ray energy selection, with Si (111) crystals used for 2600 eV photon energy, and Si (220) crystals for all higher photon energies. Samples were mounted at 10 degree x-ray incidence angle (80 degree takeoff angle).

Soft X-ray PES measurements were performed using the ambient pressure PES (APPES) endstation at the 23-ID-2 (IOS). The samples were mounted on a sample holder by carbon tape and all measurements were performed in UHV at room temperature. The endstation has a base pressure of 3x10$^{-7}$ Pa and it is equipped with a SPECS PHOIBOS 150 NAP analyzer. The incidence angle of the X-ray beam and the photoemission angle were 50 degrees and 20 degrees from the sample normal, respectively. Survey spectra were measured at a photon energy of 1260 eV while Nb 3d spectra were measured at 420 eV and 1260 eV. All spectra were measured with a pass energy of 10 eV.

RIXS measurements were performed using the Soft Inelastic X-ray (SIX) 2-ID beamline \cite{Dvorak2016,Jarrige2018}. Varied line spacing (VLS) gratings with line spacings of 500 mm$^{-1}$ and 1250 mm$^{-1}$ were respectively used on the beamline and spectrometer which, coupled with a 20 $\mu$m exit slit, provided a combined resolution of 32 meV. The spectra were measured with $\sigma$-incident polarization, where the polarization is perpendicular to the scattering plane. The incident angle was set to 10 degrees to enhance the surface contribution, and the scattering angle was set to 90 degrees.

Transmission electron microscopy work (bright field imaging, HAADF-STEM imaging, and STEM-EELS) was performed at the Center for Functional Nanomaterials (CFN) in BNL using an FEI (Thermo-Fisher Scientific) Talos F200X at an accelerating voltage of 200 kV and Gatan Enfinium EELS. TEM samples were prepared with a focused ion beam method using FEI Helios NanoLab DualBeam, via the standard in-situ lift-out method.

Temperature dependence of the resistivity of the Nb films, from 295 K down to 4 K, was performed in a Janis (model ST-3T) micromanipulated probe station at the CFN. The pressure during measurement was kept in the low 10$^{-4}$ Pa range. The Nb films had an approximate length and width of 7 mm and 3 mm, respectively. Contacts to the films were established with 25 $\mu$m Al wires in a 4-point-measurement configuration. Silver paint was used to expand the current leads across the width of the films to achieve homogeneous longitudinal current flow. Current-voltage curves, from which the resistance of the films was extracted, were measured with three Keithley source-meter units (model 2636A) as current sources (one for each Nb film) and a Keithley nanovoltmeter (model 2182A).

AFM measurements were performed at the CFN at room temperature using a commercial Park NX20-AFM instrument in tapping mode. Standard Si cantilevers with nominal resonant frequency of 330 kHz were used for scanning topography. The XEI software was used for image analysis.

\subsection*{Niobium Deposition}
\subsubsection*{Sputtered Niobium}
Niobium was deposited on 530 $\mu$m thick C-plane sapphire from CrysTec GmbH Kristalltechnologie. Prior to deposition, the sapphire was sonicated for 2 minutes each in toluene, acetone, methanol, and isopropanol, then rinsed in deionized water and blow-dried with nitrogen. The wafer was then loaded into an AJA hybrid deposition system, where it was heated to 350$^\circ$C for 20 minutes, then cooled for 20 minutes. Sputter deposition was performed at a pressure of 1.1 Pa, a power setpoint of 300 W, and a shutter delay of 60 seconds after reaching full power. The deposition was run for 252 seconds, resulting in a film thickness of 185 nm.

\subsubsection*{HiPIMS Niobium}
Niobium was deposited on 530 $\mu$m thick C-plane sapphire from CrysTec GmbH Kristalltechnologie. At Princeton University, the sapphire wafers were sonicated for 2 minutes each in toluene, acetone, methanol, and isopropanol, then rinsed in deionized water and blow-dried with nitrogen. Then, the substrates were sent to Angstrom Engineering for HiPIMS deposition of Nb. Detailed deposition parameters and photographs are presented in Supplementary Table S1 and Supplementary Figure S1. The primary changes in deposition parameters between the normal and optimized processes were substrate geometry and peak current. We hypothesize that a face-on rather than a confocal substrate geometry and higher peak currents during the HiPIMS pulse lead to higher critical temperatures. These two alterations necessitated the other process changes. In particular, the fixture fabricated by Angstrom for the face-on geometry did not allow for substrate rotation. The deposited thicknesses were 182 nm and 135 nm for HiPIMS normal and HiPIMS optimized, respectively.

\section*{Disclaimer}
Certain commercial equipment, instruments, or materials are identified in this paper in order to specify the experimental procedure adequately. Such identification is not intended to imply recommendation or endorsement by the National Institute of Standards and Technology, nor is it intended to imply that the materials or equipment identified are necessarily the best available for the purpose. Official contribution of the National Institute of Standards and Technology; not subject to copyright in the United States.

\bibliography{sample}

\section*{Acknowledgements}
This research used the SST-2, IOS, and SIX beamlines of the National Synchrotron Light Source II, and the electron microscopy facilities of the Center for Functional Nanomaterials, U.S. Department of Energy (DOE) Office of Science User Facilities operated for the DOE Office of Science by Brookhaven National Laboratory under Contract No. DE-SC0012704. Qubits were fabricated in the Princeton Institute for the Science and Technology of Materials (PRISM) cleanroom and the Quantum Device Nanofabrication Laboratory at Princeton University. A.P. acknowledges the National Science Foundation Graduate Research Fellowship, B.J. acknowledges the Humboldt Foundation, A.P.M.P. acknowledges the National Defense Science and Engineering Graduate Fellowship, and all Princeton authors acknowledge the Materials Research Science and Engineering Center (MRSEC) Grant No. DMR-1420541 and the Army Research Office Grant No. W911NF-1910016.

\section*{Competing interests}
The authors declare no competing interests. 

\section*{Author contributions statement}
I.J. and A.A.H. designed research, A.P., C.W., S.H., I.J., B.J., A.P.M.P., I.W., A.H., V.B., J.P., A.B., M.M., P.R., F.C., K.K., X.T. performed research, A.P., C.W., S.H., I.J., M.H. analyzed data, M.H. advised, A.P., I.J., M.H. wrote the main text of the manuscript. All authors reviewed the manuscript.

\end{document}


\maketitle

\section*{Qubit Fabrication}

Qubits were fabricated on each film type using the same process, except as noted. After Nb film deposition, photoresist was spun on the wafer to protect the films from dicing. The films were diced into one-inch square chips, on which photolithography was performed to define the coplanar waveguide resonator and qubit capacitor pads. The protective resist was stripped in Microposit 1165. Then, the chip was sonicated for two minutes each in toluene, acetone, methanol, and isopropanol prior to spinning photoresist AZ1505. The resist was patterned in a Heidelberg DWL66+ into a grid of nine devices, each with inputs, outputs, a resonator, and capacitor pads. To develop the resist, the chip was gently swirled in AZ300MIF for one minute, then held under running water for one minute. The exposed niobium was then descumed and etched in a PlasmaTherm APEX SLR metals etcher in an oxygen plasma and fluorine chemistry, respectively. 

Manhattan Josephson junctions were patterned using electron beam lithography. After stripping photoresist, the chip was again sonicated for two minutes each in toluene, acetone, methanol, and isopropanol. Then, a bilayer of MMA (8.5) MAA and 950 PMMA A2 (Microchem) was spun. The MMA was baked for two minutes, and the bilayer was baked for 30 minutes after spinning PMMA. A 40 nm aluminum anticharge layer was deposited using electron-beam evaporation in a Plassys MEB 550S. Then, the chip was diced into individual 7 mm by 7 mm devices. Next, the resist was patterned in an Elionix ELS-F125 electron beam lithography system: lower doses only exposed the MMA, leading to an “undercut” geometry. The anticharge layer was removed by placing the device in AZ300MIF for 4 minutes and DI water for one minute. The device was swirled in 1:3 MIBK:IPA for 50 s to develop the resist. Manhattan junctions were double-angle evaporated in a Plassys MEB 550S with a 20 nm bottom Al layer, oxidation for 15 minutes at $2 \times 10^4$ Pa, a 70 nm top Al layer, and a final oxidation for 20 minutes at $4 \times 10^3$  Pa.  The resist was lifted off in Remover PG at 80$^\circ$C for at least three hours. To complete liftoff, the device was sonicated in Remover RG for four seconds and sprayed with IPA.

The qubits were fabricated using the same CAD files, and were thus designed to have the same qubit and cavity frequencies. The measured qubit and cavity frequencies are presented in Supplementary Table \ref{table:qubit}.

Our study presents three rounds of devices, where each round had consistent fabrication, packaging, and shielding between all types of Nb film. After the first round, we improved the following procedures: (1) we used a new supply of MMA and PMMA electron-beam lithography resists, and (2) we ensured and experimentally confirmed improved thermalization between the device and the mixing chamber of the dilution refrigerator. We believe that these two changes led to overall improvements in qubit performance.

All materials characterization presented in this work (PES, HAXPES, RIXS, HAADF-STEM, TEM-EELS, AFM, and resistivity) was performed on films which came from the same one-inch square chip as the qubits for a given film type. The films used in the materials characterization underwent the same chemical processes as the qubits, described above.

\section*{Device Interfacing and Packaging}

\begin{figure}[ht]
\centering
\renewcommand{\thefigure}{S2}
\includegraphics[scale=1]{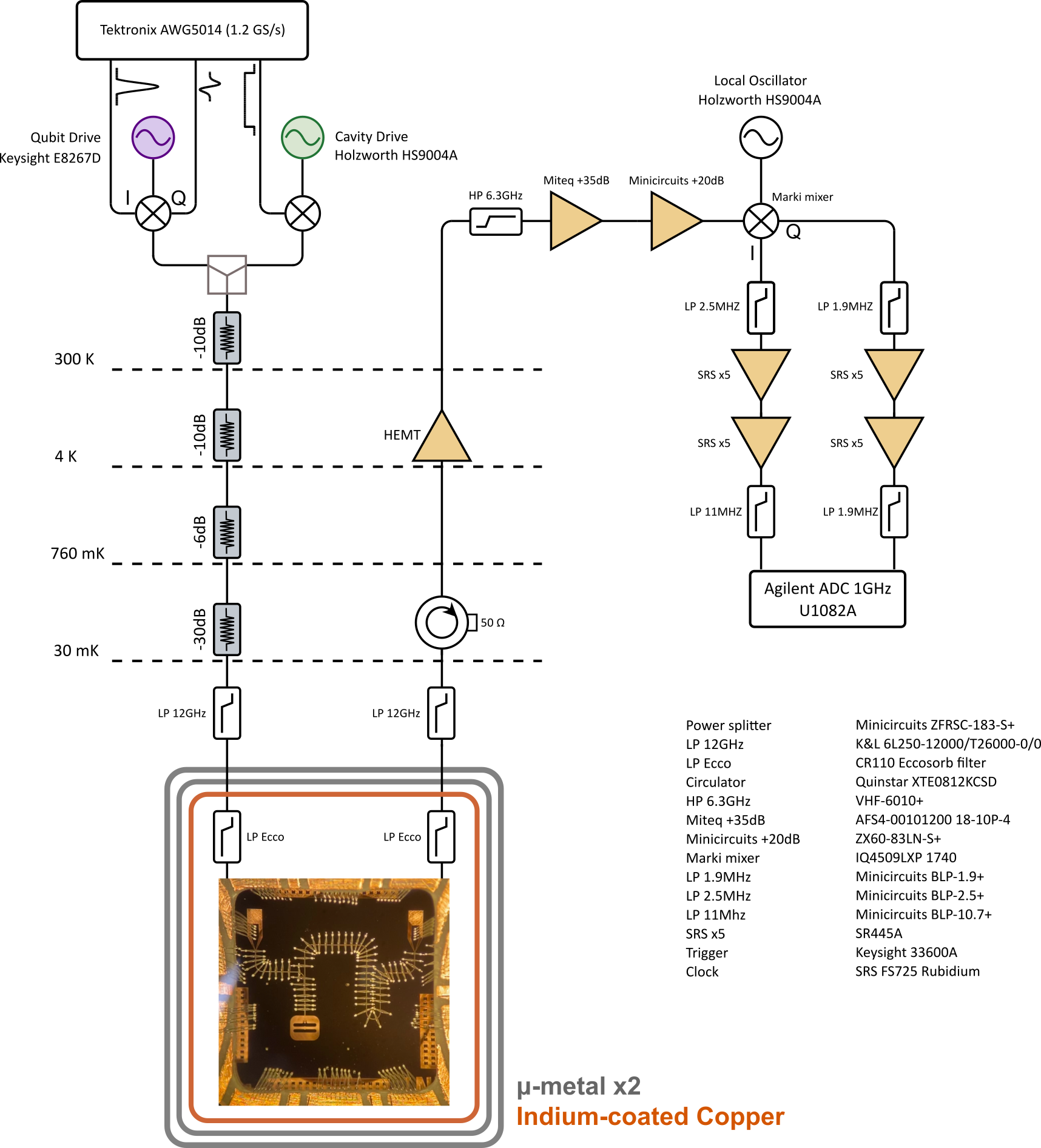}
\caption{Measurement setup and shielding. Optical image of the packaged device shows the qubit (small box in bottom-left corner) coupled to one end of a coplanar waveguide resonator. Input and output lines are capacitively coupled to the resonator, allowing for transmission measurements through the resonator and qubit drive pulses.}
\label{fig:measurement}
\end{figure}

Each device had a coplanar waveguide resonator coupled at one end to a qubit. The input line was coupled to the opposite end of the resonator, and the output line was coupled near the qubit end of the resonator, where the voltage is maximal. Devices were tightly sandwiched between a high-resistivity copper mount and a printed circuit board. Wirebonds connected the PCB to the input line, output line, and ground plane. A high-resistivity copper cap was placed atop the PCB where the device was exposed. The copper mount was thermally sunk to the mixing chamber of a dilution refrigerator at 30 mK with a high-resistivity copper rod. The entire device package was shielded with one inner layer of indium-coated copper and two outer layers of $\mu$-metal. Each $\mu$-metal shield was covered in aluminized mylar on the outside, and the inside of the inner copper shield was coated with stycast and grains of SiC. Refer to Supplementary Figure \ref{fig:measurement} for an interfacing and packaging schematic.

\section*{Measurement and Control}

\begin{figure}[ht]
\centering
\renewcommand{\thefigure}{S3}
\includegraphics[scale=0.47]{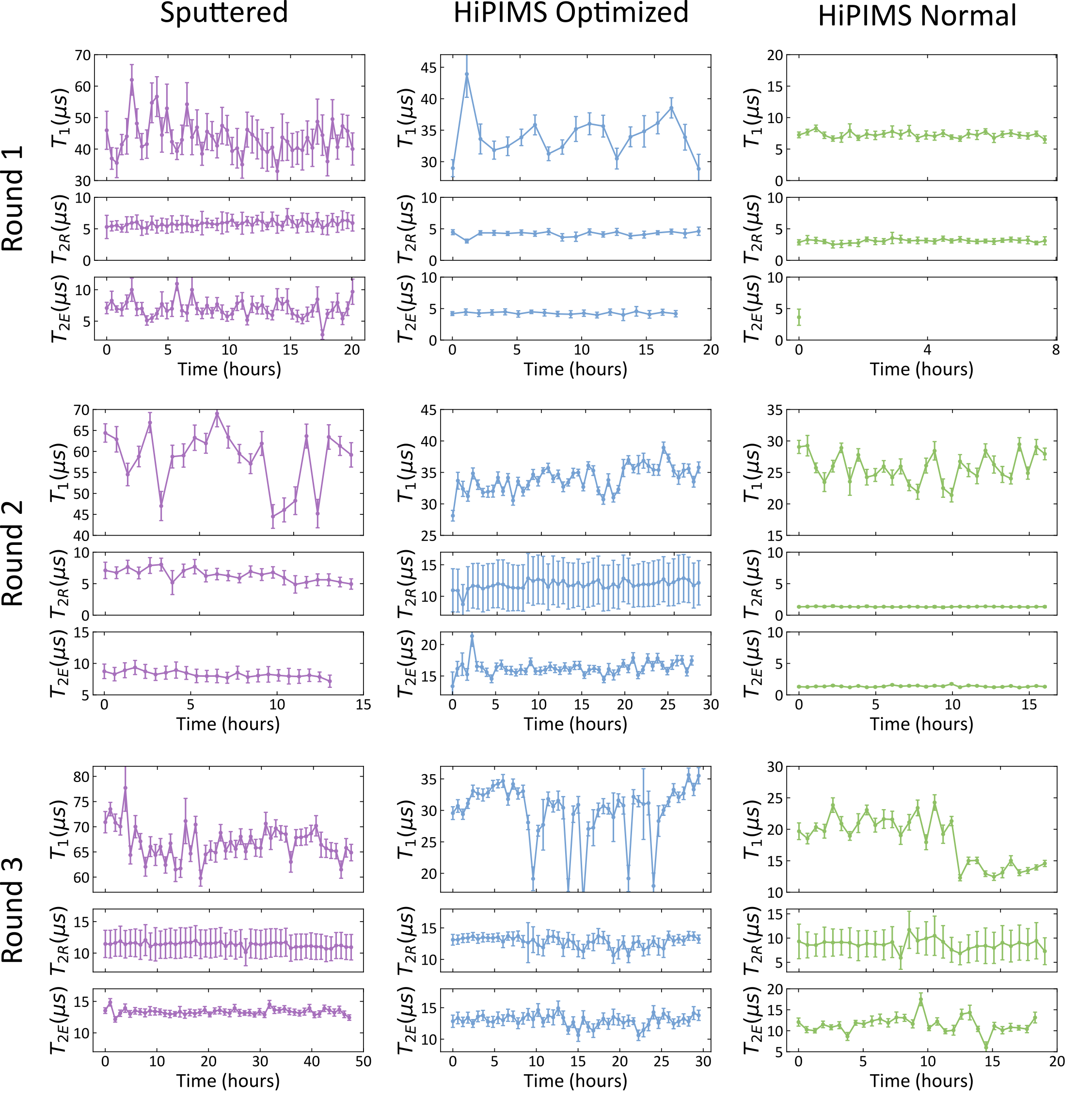}
\caption{Time traces for relaxation ($T_1$), Ramsey ($T_{2R}$), and echo ($T_{2E}$) measurements for all devices. Error bars on each measurement indicate fit error.}
\label{fig:timetraces}
\end{figure}

We performed time-domain measurements to characterize qubit performance. Relaxation ($T_1$) measurements were performed by exciting the qubit with a Gaussian $\pi$-pulse at the qubit frequency, then reading out the qubit state after delay time $\tau$. Ramsey ($T_{2\mathrm{R}}$) measurements were performed with two Gaussian $\frac{\pi}{2}$ pulses at the qubit frequency separated by delay time $\tau$. The phase of the second pulse was given by $2 \pi f_{\mathrm{fringe}} \tau$, where $f_{\mathrm{fringe}}$ is the desired fringe frequency to aid fits. Readout occurred immediately after the second pulse. Echo ($T_{2\mathrm{E}}$) measurements involved, in addition to the Ramsey pulse scheme, an additional $\pi$ pulse halfway between the two $\frac{\pi}{2}$ pulses, in phase with the first pulse. In cases where $T_{2\mathrm{E}}$ was fit using exponential decay rather than exponentially decaying fringes, the fringe frequency $f_{\mathrm{fringe}}$ was set to zero. Measured relaxation, Ramsey, and echo times are presented in Supplementary Table \ref{table:qubit} and Supplementary Figure \ref{fig:timetraces}. The reported relaxation and coherence times for a single device are the average of 19-50 sequential measurements. 

The control and measurement setup is shown in Supplementary Figure \ref{fig:measurement}. Gaussian pulses at the qubit frequency were synthesized via IQ mixing with a Tektronix AWG5014 and a Keysight E8267D Vector Signal Generator. Square pulses at the cavity frequency for readout were synthesized using an on-off gate from the Tektronix to switch the Holzworth HS9004A phase-coherent RF synthesizer. The qubit and cavity signals were combined, sent into the fridge, and thermalized at subsequent stages with attenuators. For both the input and output lines, we used KNL low-pass filters outside the shields and CR110 Eccosorb filters inside the shields to attenuate high-frequency radiation. 

The output was sent through a Quinstar four-port circulator mounted to the mixing chamber with two inputs terminated with 50 ohms, then sent to a high electron mobility transistor (HEMT) low-noise amplifier mounted to the 4K stage. At room temperature, a high-pass filter at 6.3 GHz filtered out the qubit pulse to avoid saturating the subsequent amplifiers (Miteq and Minicircuits). The signal was mixed in homodyne with a local oscillator, which was produced by the same Holzworth as the readout tone and was thus phase-locked with the signal. The resulting I and Q DC signals were cleaned up with low-pass filters and amplified twice each using an SRS SR445A, then sent directly to an Agilent ADC U1082A. 

The Tektronix AWG and ADC were triggered with a Keysight 33600A Series waveform generator using a square pulse at 50 $\%$ duty cycle. Each point in a single $T_1$ trace was the average of 4000 - 20000 trigger cycles. For each trigger cycle, the ADC read the I and Q signals for 8.2 $\mathrm{\mu s}$ with a sample rate of 1 GHz. Finally, all electronic equipment was clocked with an SRS F2725 Rubidium Frequency Standard. 

\section*{Disclaimer}
Certain commercial equipment, instruments, or materials are identified in this paper in order to specify the experimental procedure adequately. Such identification is not intended to imply recommendation or endorsement by the National Institute of Standards and Technology, nor is it intended to imply that the materials or equipment identified are necessarily the best available for the purpose.

\pagebreak

\begin{table}
\centering
\renewcommand{\thetable}{S1}
\begin{tabular}{ |l| c c| }
 \hline
 Parameter & HiPIMS Normal & HiPIMS Optimized \\ 
 \hline
 Initial Pressure & 4.0 Pa & 1.3 Pa \\  
 Substrate Geometry & Confocal & Face-On \\
 Substrate Rotation & 10 rpm & none \\  
 Source-substrate Distance & 4.5 in & 3.5 in \\
 Process Pressure & 4000 Pa & 1300 Pa \\ 
 Pulse length & 50 $\upmu$s & 75 $\upmu$s \\
 Pulse repeat rate & 250 Hz & 250 Hz \\
 Voltage & 568 V & 640 V \\  
 Current & 678 mA & 1565 mA \\
 Power & 385 W & 1001 W \\  
 Peak Current & 78 A & 241 A \\
 Gas Flow (Ar) & 20 sccm & 20 sccm \\
 Deposition time & 2667 s & 230 s \\
 Total thickness & 182 nm & 135 nm \\
 \hline
\end{tabular}
\caption{Deposition parameters for HiPIMS normal and HiPIMS optimized films. }
\label{table:HiPIMS}
\end{table}

\begin{figure}[ht]
\centering
\renewcommand{\thefigure}{S1}
\includegraphics[scale=1]{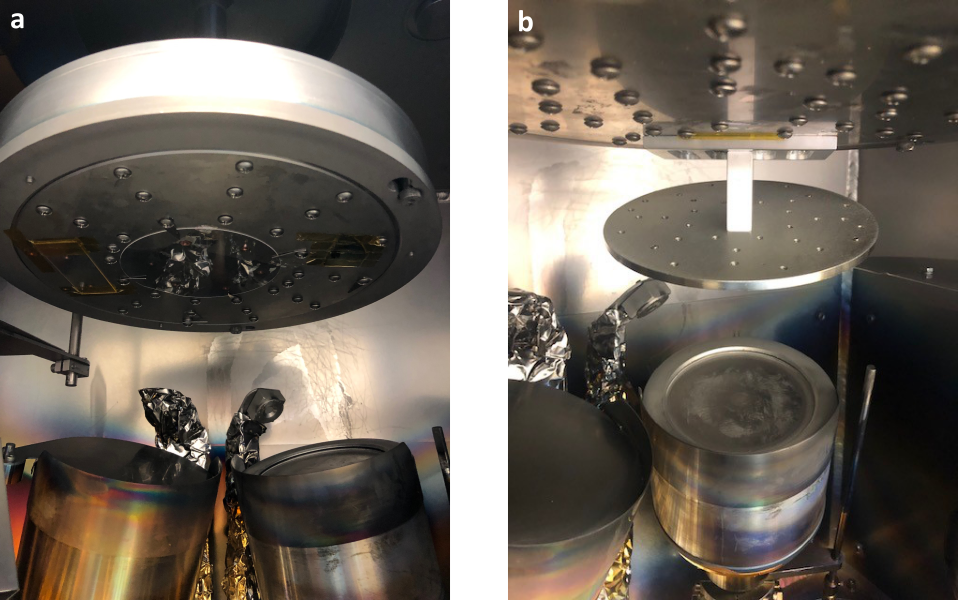}
\caption{Photographs of substrate geometries for HiPIMS deposition. (a) Confocal geometry for the HiPIMS normal deposition. Such a geometry is commonly used in sputtering deposition. (b) Face-on geometry for the HiPIMS optimized deposition. An angled chuck was fabricated such that the faces of the substrate and target would be parallel.}
\label{fig:HiPIMSpics}
\end{figure}

\setlength{\tabcolsep}{9pt}

\begin{table}
\centering
\renewcommand{\thetable}{S2}
\begin{tabular}{ |l| l l l l c m{1.6cm} m{1.7cm}| }
 \hline
 Device & T$_1$ ($\upmu$s) & T$_{2\mathrm{R}}$ ($\upmu$s) & T$_{2\mathrm{E}}$ ($\upmu$s) & T$_{2\mathrm{E}}$ fit & N & Qubit Freq. (GHz) & Cavity Freq. (GHz)\\ 
 \hline
 Sputtered \#1 & 44 $\pm$ 6 & 5.8 $\pm$ 0.4 & 7 $\pm$ 1 & Exp & 50 & 4.534 & 7.351 \\  
 HiPIMS Opt \#1 & 34 $\pm$ 4 & 4.2 $\pm$ 0.4 & 4.3 $\pm$ 0.2 & ExpCos & 19 & 4.552 & 7.249 \\ 
 HiPIMS Norm \#1 & 7.3 $\pm$ 0.4 & 3.1 $\pm$ 0.2 & 4 $\pm$ 1$^\dagger$ & ExpCos & 30 & 4.444 & 7.187 \\ 
 \hline
 Sputtered \#2 & 58 $\pm$ 7 & 6.4 $\pm$ 1.0 & 8.3 $\pm$ 0.5 & ExpCos & 23 & 4.955 & 7.331 \\
 HiPIMS Opt \#2 & 34 $\pm$ 2 & 11.9 $\pm$ 0.7 & 16.3 $\pm$ 0.7 & Exp & 50 & 4.515 & 7.227 \\  
 HiPIMS Norm \#2 & 26 $\pm$ 2 & 1.33 $\pm$ 0.04 & 1.4 $\pm$ 0.1 & Exp & 30 & 5.300 & 7.201 \\ 
 \hline
 Sputtered \#3 & 67 $\pm$ 3 & 11.4 $\pm$ 0.4 & 13.3 $\pm$ 0.5 & Exp & 50 & 5.008 & 7.333 \\ 
 HiPIMS Opt \#3 & 30 $\pm$ 5 & 12.8 $\pm$ 0.8 & 13.0 $\pm$ 0.9 & Exp & 50 & 4.762 & 7.229 \\
 HiPIMS Norm \#3 & 18 $\pm$ 4 & 9 $\pm$ 1 & 12 $\pm$ 2 & Exp & 30 & 4.660 & 7.182 \\  
 \hline
\end{tabular}
\caption{Qubit performance and measured device parameters. Each coherence time is the average and standard deviation of N sequential measurements, all of which are shown in Supplementary Figure \ref{fig:timetraces}. In the $T_{2\mathrm{E}}$ fit column, "Exp" denotes $f_{\mathrm{fringe}} = 0$ with an exponential decay fit, and and "ExpCos" denotes a non-zero $f_{\mathrm{fringe}}$ with a exponentially decaying sinusoidal fit. $\dagger$ indicates that only one measurement was taken, and the error is the uncertainty of the fit.}
\label{table:qubit}
\end{table}

\begin{figure}[ht]
\centering
\renewcommand{\thefigure}{S4}
\includegraphics[scale=0.7]{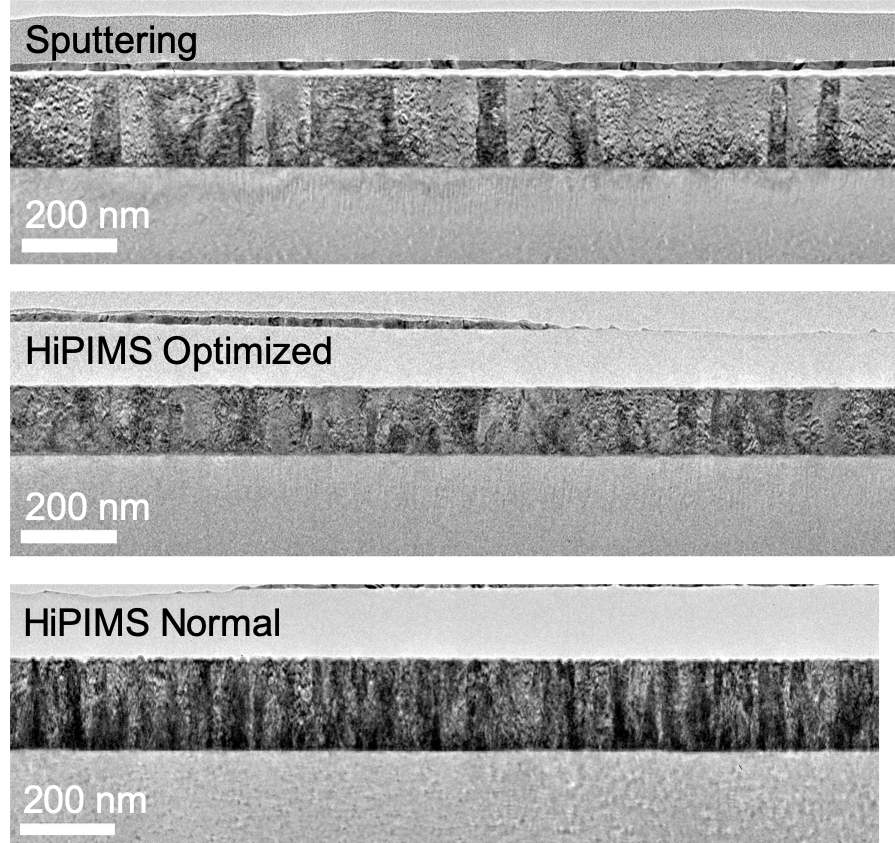}
\caption{Low-magnification TEM bright-field images of the three films, showing the largely columnar profile of the grains.}
\label{fig:TEMphotos}
\end{figure}

\begin{figure}[ht]
\centering
\renewcommand{\thefigure}{S5}
\includegraphics[width=\linewidth]{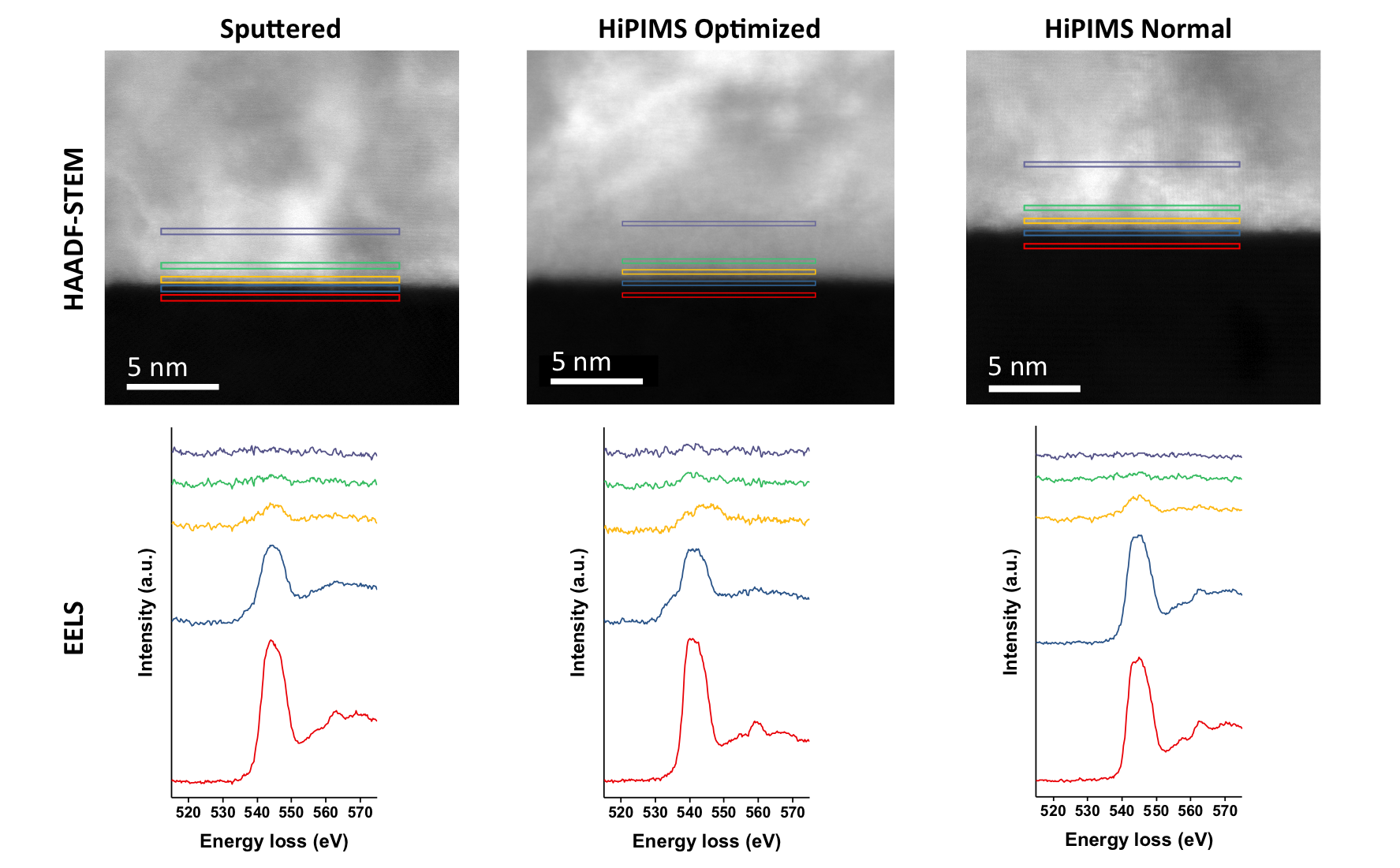}
\caption{Structural and chemical imaging of the substrate-metal interface for the sputtered (left), HiPIMS optimized (middle), and HiPIMS normal (right) films. The top row shows high-angle annular dark-field scanning transmission electron spectroscopy (HAADF-STEM) measurements at cross sections. The second row shows O-$K$-edge electron-energy loss spectroscopy (EELS) spectra measured at the locations indicated on the HAADF-STEM images. The horizontal bars on the HAADF-STEM images indicate a range of locations that were averaged over to obtain the EELS spectra shown. At this level, we observe a chemically sharp transition from the oxidized sapphire substrate to the Nb film, which shows little oxidation. The amount and locations of oxidation are similar across films.}
\label{fig:TEMphotos2}
\end{figure}

\begin{figure}[ht]
\centering
\renewcommand{\thefigure}{S6}
\includegraphics[width=\linewidth]{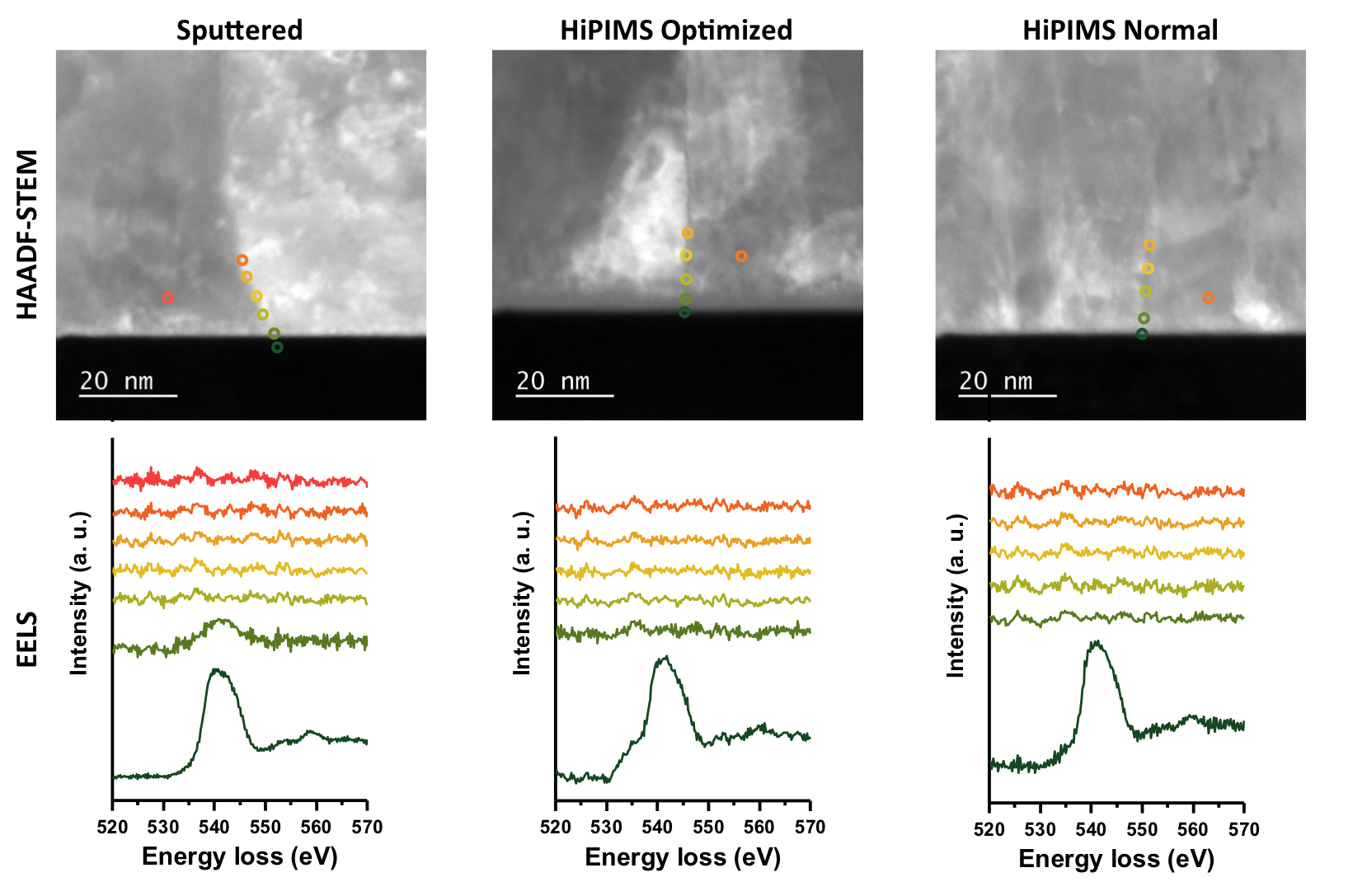}
\caption{Structural and chemical imaging of grain boundaries at the substrate-metal interface for the the sputtered (left), HiPIMS optimized (middle), and HiPIMS normal (right) films. The top row shows HAADF-STEM measurements at cross sections. The second row shows O-$K$-edge EELS spectra measured at the locations indicated on the HAADF-STEM images. The large peak in the darkest green spectra corresponds to oxidation in the sapphire substrate. The reddest spectrum for each film shows very little oxidation away from the grain boundary. Spectra taken along the grain boundary do not show stronger oxidation than the reddest spectrum for any film, indicating that the grain boundaries at this interface are consistently unoxidized at this level. }
\label{fig:TEMphotos3}
\end{figure}

\clearpage